\documentclass[12pt,cite,a4paper]{article}
\usepackage{graphicx}
\usepackage{cite}
\usepackage{amsmath}
\usepackage{amsfonts}
\usepackage{amssymb}
\begin{document}
\hfill\hbox{RIKEN-TH-82}

\hfill\hbox{PTA/06-23}

\bigskip\ 

\begin{center}
{\Large \textbf{Higher Equations of Motion in $\mathbf{N=1}$ SUSY Liouville
Field Theory}}

\vspace{1.5cm}

{\large A.Belavin}

\vspace{0.2cm}

L.D.Landau Institute for Theoretical Physics RAS,

142432 Chernogolovka, Russia

\vspace{0.2cm}

and

\vspace{0.2cm}

{\large Al.Zamolodchikov}\footnote{On leave of absence from Institute of
Theoretical and Experimental Physics, B.Cheremushkinskaya 25, 117259 Moscow, Russia.}

\vspace{0.2cm}

Laboratoire de Physique Th\'eorique et Astroparticules, UMR-5207 CNRS-UM2,

Universit\'e Montpellier II, Pl.E.Bataillon, 34095 Montpellier, France

\vspace{0.2cm}

and

\vspace{0.2cm}

Service de Physique Th\'eorique, CNRS - URA 2306,

C.E.A. - Saclay, F-91191, Gif-sur-Yvette, France
\end{center}

\vspace{1.0cm}

\textbf{Abstract}

Similarly to the ordinary bosonic Liouville field theory, in its $N=1$
supersymmetric version an infinite set of operator valued relations, the
``higher equations of motions'', holds. Equations are in one to one
correspondence with the singular representations of the super Virasoro algebra
and enumerated by a couple of natural numbers $(m,n)$. We demonstrate
explicitly these equations in the classical case, where the equations of type
$(1,n)$ survive and can be interpreted directly as relations for classical
fields. General form of the higher equations of motion is established in the
quantum case, both for the Neveu-Schwarz and Ramond series.

\vspace{0.5cm}

\textbf{Motivations.} In ref. \cite{higher} it has been shown that in the
Liouville field theory (LFT) an infinite set of relations holds for quantum
operators. They are paramete\-rized by pairs of positive integers $(m,n)$ and
called conventionally the ``higher equations of mo\-tion'' (HEM), because the
first one $(1,1)$ coincides with the usual Liouville equation of motion. These
equations relate different basic LFT primary fields $V_{a}(x)$ (operators
$V_{a}$ can be thought of as regularized version of the exponential
$\exp(2a\phi)$ of the basic Liouville field $\phi$). The equations are
``derived'' on the basis of two conjectures. First one is the vanishing of all
singular vectors in the representations built on the exponential fields. This
is a natural continuation of the easily verified relations in the classical
LFT, and also a mandatory requirement imposed on LFT by the unitarity. The
second conjecture is much less justified and states basically that the set of
exponential fields $\left\{  V_{a}\right\}  $ (with complex $a$ allowed)
covers in some sense the whole variety of primary fields in LFT. One of the
open problems here is the concept of the space of local fields and its
completeness. In LFT, unlike the more familiar rational conformal field
theories \cite{BPZ}, the operator state correspondence doesn't hold literally,
thus making difficult a straightforward inheritance of the completeness from
the unitary space of physical states. Because of this conceptual problem the
status of the second conjecture in unclear and waits for a mathematically
correct formulation. For the moment we simply take for granted that the space
of primary local fields, the norms and completeness regardless, is spanned by
an appropriate subset of $\left\{  V_{a}\right\}  $.

Similar operator valued relations have been observed also in $SL(2,R)_{k}$
WZNW model \cite{WZNW}. Further study of these and the Liouville related HEM's
can be found in ref.\cite{Matone}.

Higher equations turned to be useful in practical calculations. In particular,
in \cite{BAl} they were used to derive general four-point correlation function
in the minimal Liouville gravity (see \cite{threep} for the dictionary) with
one degenerate matter field. It's very likely that HEM's are potentially
important in the general program of explicit construction of the complete set
of correlation functions in the minimal Liouville gravity.

It is the purpose of this note to reveal a similar set of higher equations in
the supersymmetric Liouville field theory (SLFT). The $N=1$ SUSY version of
LFT \cite{Polyakov} is known to be closest partner of the bosonic theory,
having very similar properties. In particular, the existence of a
supersymmetric version of the HEM's is naturally expected.

\textbf{Higher equations in classical SLFT.} We begin with the classical
equations of motion in the $N=1$ SLFT\cite{Arvis, DHoker}
\begin{align}
\bar\partial\psi_{\text{c}} &  =iM\bar\psi_{\text{c}}e^{\varphi}\nonumber\\
\partial\bar\psi_{\text{c}} &  =-iM\psi_{\text{c}}e^{\varphi}\label{csliouv}\\
\partial\bar\partial\varphi &  =iM\bar\psi_{\text{c}}\psi_{\text{c}}%
e^{\varphi}+M^{2}e^{2\varphi}\nonumber
\end{align}
where $\varphi$ is the boson and $(\psi_{\text{c}},\bar\psi_{\text{c}})$ the
Majorana fermion components\footnote{index ``c'' is attached to the fields
$\psi_{\text{c}}$, and also to the classical supercurrent $S_{\text{c}}$ and
to the stress tensor $T_{\text{c}}$, to distinguish them from differently
normalized quantum fields, which appear in the subsequent sections.}. As in
\cite{higher} we use complex 2D coordinates $z=x+iy$ and $\bar z=x-iy$
($\partial=\partial_{z}$ and $\bar\partial=\partial_{\bar z}$) and introduce a
redundant parameter $M$ for the sake of convenience. Equations (\ref{csliouv})
can be obtained as the extremum conditions for the following classical
Lagrangian density
\begin{equation}
\mathcal{L}_{\text{cl}}=\partial\varphi\bar\partial\varphi+\psi_{\text{c}}%
\bar\partial\psi_{\text{c}}+\bar\psi_{\text{c}}\partial\bar\psi_{\text{c}%
}+2iM\bar\psi_{\text{c}}\psi_{\text{c}}e^{\varphi}+M^{2}e^{2\varphi
}\ .\label{Scl}%
\end{equation}
The classical (as well as the quantum) SLFT has been introduced and studied in
\cite{Arvis, DHoker, Babelon} shortly after it appeared in the string context
in \cite{Polyakov}. Here we recapitulate only those properties of the
classical theory which will be of further use.

The superconformal invariance at the classical level is equivalent to the
statement that the components of the classical supercurrent
\begin{align}
S_{\text{c}} &  =-i\psi_{\text{c}}\partial\varphi+i\partial\psi_{\text{c}%
}\label{Sc}\\
\ \ \bar S_{\text{c}} &  =-i\bar\psi_{\text{c}}\bar\partial\varphi
+i\bar\partial\bar\psi_{\text{c}}\nonumber
\end{align}
are holomorphic $\bar\partial S_{\text{c}}=0$ and antiholomorphic
$\partial\bar S_{\text{c}}=0$ respectively. These relations, as well as
similar statements $\bar\partial T_{\text{c}}=\partial\bar T_{\text{c}}=0$
about the stress tensor components
\begin{align}
T_{\text{c}} &  =-\frac12(\partial\varphi)^{2}+\frac12\partial^{2}%
\varphi+\frac12\partial\psi_{\text{c}}\psi_{\text{c}}\label{Tc}\\
\bar T_{\text{c}} &  =-\frac12(\bar\partial\varphi)^{2}+\frac12\bar
\partial^{2}\varphi+\frac12\bar\partial\bar\psi_{\text{c}}\bar\psi_{\text{c}%
}\nonumber
\end{align}
are easily verified to be consequences of the equations (\ref{csliouv}). To
fully describe the supersymmetry we need also classical generators $G$ and
$\bar G$, the right and left supercharges. These operators act on the
classical fields and satisfy the standard relations
\begin{equation}
G^{2}=\partial\;;\ \bar G^{2}=\bar\partial\;;\ \left\{  G,\bar G\right\}
=0\ .\label{cSUSY}%
\end{equation}
Their action on the fundamental components $\varphi$ and $(\psi_{\text{c}%
},\bar\psi_{\text{c}})$ is
\begin{equation}
G\varphi=i\psi_{\text{c}}\;;\;\;\;\bar G\varphi=i\bar\psi_{\text{c}%
}\ .\label{Qphi}%
\end{equation}
The action of the supercharges on a general exponential field is a direct
consequence of (\ref{Qphi})
\begin{equation}
Ge^{\sigma\varphi}=i\sigma\psi_{\text{c}}e^{\varphi}\;;\ \ \bar Ge^{\sigma
\varphi}=i\sigma\bar\psi_{\text{c}}e^{\varphi}\ .\label{Qexp}%
\end{equation}

All three equations (\ref{csliouv}) follow from the statement
\begin{equation}
\bar GG\varphi=iMe^{\varphi}\label{QL}%
\end{equation}
and the algebra (\ref{cSUSY}). This gives, after a simple reckoning
\begin{equation}
\bar GGe^{\sigma\varphi}=iM\sigma e^{(1+\sigma)\varphi}-\sigma^{2}\bar
\psi_{\text{c}}\psi_{\text{c}}e^{\sigma\varphi}\;,\label{QQexp}%
\end{equation}
a relation useful in the subsequent calculations.

The classical $D$-operators form an infinite series $D_{2k-1}^{\text{(c)}}$,
$k=1,2,\ldots$. Few first representatives read
\begin{align}
D_{1}^{\text{(c)}} &  =G\nonumber\\
D_{3}^{\text{(c)}} &  =G\partial+S_{\text{c}}\nonumber\\
D_{5}^{\text{(c)}} &  =G\partial^{2}+2T_{\text{c}}G+3S_{\text{c}}%
\partial+2\partial S_{\text{c}}\label{Dn}\\
D_{7}^{\text{(c)}} &  =G\partial^{3}+8T_{\text{c}}G\partial+4\partial
T_{\text{c}}G+18T_{\text{c}}S_{\text{c}}+6S_{\text{c}}\partial^{2}+8\partial
S_{\text{c}}\partial+3\partial^{2}S_{\text{c}}\nonumber\\
D_{9}^{\text{(c)}} &  =G\partial^{4}+20T_{\text{c}}G\partial^{2}+20\partial
T_{\text{c}}G\partial+6\partial^{2}T_{\text{c}}G+36T_{\text{c}}^{2}%
G+110T_{\text{c}}S_{\text{c}}\partial\nonumber\\
&  \ +56\partial T_{\text{c}}S_{\text{c}}+72T_{\text{c}}\partial S_{\text{c}%
}+10S_{\text{c}}\partial^{3}+20\partial S_{\text{c}}\partial^{2}%
+15\partial^{2}S_{\text{c}}\partial+4\partial^{3}S_{\text{c}}+\partial
S_{\text{c}}S_{\text{c}}G\nonumber\\
&  \ \ldots\nonumber
\end{align}
There is, of course, the identical series of the ``left'' operators $\bar
D_{2k-1}^{\text{(c)}}$. One only needs to combine the SUSY algebra
(\ref{cSUSY}) with the definitions of the supercurrent (\ref{Sc}) and the
stress tensor (\ref{Tc}) to verify the identities
\begin{equation}
D_{2k-1}^{\text{(c)}}e^{-(k-1)\varphi}=\bar D_{2k-1}^{\text{(c)}%
}e^{-(k-1)\varphi}=0\label{cnull}%
\end{equation}

Direct calculation with a help of the equations of motion (\ref{csliouv})
gives
\begin{align}
\bar D_{1}^{\text{(c)}}D_{1}^{\text{(c)}}\varphi &  =iMe^{\varphi}\nonumber\\
\bar D_{3}^{\text{(c)}}D_{3}^{\text{(c)}}\varphi e^{-\varphi} &
=-iM^{3}e^{2\varphi}\label{todo}\\
\bar D_{5}^{\text{(c)}}D_{5}^{\text{(c)}}\varphi e^{-2\varphi} &
=4iM^{5}e^{3\varphi}\nonumber\\
\bar D_{7}^{\text{(c)}}D_{7}^{\text{(c)}}\varphi e^{-3\varphi} &
=-36iM^{7}e^{4\varphi}\;\nonumber\\
\bar D_{7}^{\text{(c)}}D_{7}^{\text{(c)}}\varphi e^{-3\varphi} &
=576iM^{9}e^{5\varphi}\;.\nonumber
\end{align}
This allows to conjecture that for general $k=1,2,\ldots$
\begin{equation}
\bar D_{2k-1}^{\text{(c)}}D_{2k-1}^{\text{(c)}}\varphi e^{(1-k)\varphi
}=i(-)^{k-1}[(k-1)!]^{2}M^{2k-1}e^{k\varphi}\label{cHEM}%
\end{equation}
We will show in the subsequent sections that this is a classical limit of a
(subset of) more general set of relations, the HEM's of the quantum SLFT.

\textbf{Quantum SLFT.} We remind very briefly the essence of the quantum SLFT
(see \cite{DK, Pogossian, Marian}). The Lagrangian density is\footnote{In this
paper we use the component form of the supersymmetric expressions,
systematically avoiding superspace notations. Presently the supefields do not
give essential economy, neither notational, nor technical.}
\begin{equation}
\mathcal{L}_{\text{SLFT}}=\frac1{8\pi}\left(  \partial_{a}\phi\right)
^{2}+\frac1{2\pi}\left(  \psi\bar\partial\psi+\bar\psi\partial\bar\psi\right)
+2i\mu b^{2}\bar\psi\psi e^{b\phi}+2\pi b^{2}\mu^{2}e^{2b\phi}\label{SL}%
\end{equation}
Here $\mu$ is a scale coupling called conventionally the cosmological constant
while $b$ is a quantum parameter, the classical limit corresponding to
$b\rightarrow0$. A convenient combination
\begin{equation}
Q=b^{-1}+b\label{Q}%
\end{equation}
is traditionally called the background charge. In the form (\ref{SL}) the
action is explicitly supersymmetric, the classical one (\ref{Scl}) being
achieved straightforwardly in the limit $b\rightarrow0$ with $b\phi
\rightarrow\varphi$, $b\psi\rightarrow\psi_{\text{c}}$, $2\pi\mu
b^{2}\rightarrow M$ and
\begin{equation}
\int\mathcal{L}_{\text{SLFT}}d^{2}x\rightarrow\frac1{2\pi b^{2}}S_{\text{cl}%
}\label{SLtocl}%
\end{equation}
For the ``perturbed CFT'' interpretation the last two terms in (\ref{SL}) are
better understood through the field
\begin{equation}
-\bar GGe^{b\phi}=b^{2}\bar\psi\psi e^{b\phi}-2i\pi\mu b^{2}e^{2b\phi
}\;,\label{Lint}%
\end{equation}
a ``top'' component of an appropriate supermultiplet.

SLFT is a superconformal field theory (SCFT), the symmetry being generated by
the holomorphic and antiholomorphic components of the supercurrent
\begin{equation}
S(z)=-i\phi\partial\psi+iQ\partial\psi\ ;\ \bar S(\bar z)=-i\phi\bar
\partial\bar\psi+iQ\partial\bar\psi\label{S}%
\end{equation}
In the classical limit they apparently turn to the fields (\ref{Sc}) as
$S\rightarrow b^{-2}S_{\text{c}}$, $\bar S\rightarrow b^{-2}\bar S_{\text{c}}%
$. In the same way the classical holomorphic and antiholomorphic stress tensor
components (\ref{Tc}) are the limits $T\rightarrow b^{-2}T_{\text{c}}$, $\bar
T\rightarrow b^{-2}\bar T_{\text{c}}$ of the holomorphic and antiholomorphic
quantum fields
\begin{align}
T(z) &  =-\frac12(\partial\phi)^{2}+\frac Q2\partial^{2}\phi+\frac
12\partial\psi\psi\label{T}\\
\bar T(\bar z) &  =-\frac12(\bar\partial\phi)^{2}+\frac Q2\bar\partial^{2}%
\phi+\frac12\bar\partial\bar\psi\bar\psi\nonumber
\end{align}
Together with the supercurrent (\ref{S}) they form the superconformal algebra
of operator product expansions
\begin{align}
S(z)S(z^{\prime}) &  =\frac{\widehat{c}}{(z-z^{\prime})^{3}}+\frac
{T(z^{\prime})}{z-z^{\prime}}+\text{reg.}\nonumber\\
T(z)S(z^{\prime}) &  =\frac{3S(z^{\prime})}{2(z-z^{\prime})^{2}}%
+\frac{\partial S(z^{\prime})}{z-z^{\prime}}+\text{reg.}\label{SCOPE}\\
T(z)T(z^{\prime}) &  =\frac{3\widehat{c}}{4(z-z^{\prime})^{4}}+\frac
{2T(z^{\prime})}{(z-z^{\prime})^{2}}+\frac{\partial T(z^{\prime})}%
{z-z^{\prime}}+\text{reg.}\nonumber
\end{align}
where the central charge is related to $b$ as
\begin{equation}
\widehat{c}=1+2Q^{2}\label{cQ}%
\end{equation}
In terms of the Laurent components
\begin{equation}
T(z)=\sum_{n}L_{n}z^{-n-2}\ ;\ \ \ S(z)=\sum_{k}G_{k}z^{-k-3/2}\label{Laurent}%
\end{equation}
(index $k\in Z+1/2$ in the Neveu-Schwarz (NS) representations and $k\in Z$ in
the Ramond (R) ones) the algebra takes the conventional form of super Virasoro
algebra ($SV$)
\begin{align}
\{G_{k},G_{l}\} &  =2L_{k+l}+\frac{\widehat{c}}2\left(  k^{2}-\frac14\right)
\delta_{k+l}\nonumber\\
\lbrack L_{n},G_{k}] &  =\left(  \frac n2-k\right)  G_{n+k}\label{SVir}\\
\lbrack L_{m},L_{n}] &  =(m-n)L_{m+n}+\frac{\widehat{c}}8(m^{3}-m)\delta
_{m+n}\nonumber
\end{align}
An identical, ``left'' algebra, is generated by the antiholomorphic components
$\bar S(\bar z)$ and $\bar T(\bar z)$ and their Laurent components $\bar
G_{k}$ and $\bar L_{n}$.

Local fields form the highest weight representaitons of the right and left
algebras $SV\otimes\overline{SV}$, both either Neveu-Schwarz or Ramond ones.
The corresponding highest weight vectors are the SCFT primary fields, denoted
$V_{a}$ for the NS representations and $R_{a}^{\pm}$ for the R
representations. Their dimensions depend on the parameter $a$ as
\begin{equation}
\Delta_{\text{NS}}(a)=\frac{a(Q-a)}2\ ;\ \ \Delta_{\text{R}}(a)=\Delta
_{\text{NS}}(a)+\frac1{16}\;,\label{DNS}%
\end{equation}
differently for the NS and R sectors. It is not a bad idea to compare the
basic SLFT fields $\phi$ and $(\psi,\bar\psi)$ with free massless boson and
Majorana fermion. In this dictionary at $\operatorname*{Re}a<Q/2$ the primary
field $V_{a}$ corresponds to the normal ordered exponential $:\exp(a\phi):$.
Another convenient parameter $\lambda=Q/2-a$ is often used instead of $a$.
Correspondingly
\begin{equation}
\Delta_{\text{NS}}(a)=\frac{(\widehat{c}-1)}{16}-\frac{\lambda^{2}%
}2\;;\;\;\Delta_{\text{R}}(a)=\frac{\widehat{c}}{16}-\frac{\lambda^{2}%
}2\,.\label{Dlambda}%
\end{equation}
In the last Ramond case there are two equally good highest weight vectors
$R_{a}^{\pm}$, forming a two dimensional representation
\begin{align}
G_{0}\binom{R_{a}^{+}}{R_{a}^{-}}  & =\left(
\begin{array}
[c]{cc}%
0 & \dfrac{(-1+i)\lambda}2\\
\dfrac{(1+i)\lambda}2 & 0
\end{array}
\right)  \binom{R_{a}^{+}}{R_{a}^{-}}\label{G0}\\
\bar G_{0}\binom{R_{a}^{+}}{R_{a}^{-}}  & =\left(
\begin{array}
[c]{cc}%
0 & \dfrac{(1+i)\lambda}2\\
\dfrac{(-1+i)\lambda}2 & 0
\end{array}
\right)  \binom{R_{a}^{+}}{R_{a}^{-}}\nonumber
\end{align}
of the Cartan subalgebra $G_{0}^{2}=\bar G_{0}^{2}=L_{0}-\widehat{c}/16$ and
$\left\{  G_{0},\bar G_{0}\right\}  =0$. In the free field language $R_{a}%
^{+}$ and $R_{a}^{-}$ can be related respectively to the fields $\sigma
:\exp(a\phi):$ and $\mu:\exp(a\phi):$, where $\sigma$ and $\mu$ are the
standard order and disorder spin fields with respect to the free fermion,
familiar from the Ising model \cite{KC}. In our further development this
doubling of the Ramond primary fields is not very important and we will often
omit the index $\pm$ near $R_{a}^{\pm}$, keeping however in mind this feature.

\textbf{Degenerate primaries.} At certain special values of the parameter $a$
the $SV$ representations are singular. This happens at \cite{Kac} $a=a_{m,n}$
(or, equivalently, at $a=Q-a_{m,n}$), where $a_{m,n}=-\lambda_{m-1,n-1}$ and
$(m,n)$ is a pair of positive integers. We introduced a convenient notaiton
\begin{equation}
\lambda_{m,n}=\frac{mb^{-1}+nb}2\label{lmn}%
\end{equation}
In general at $a=a_{m,n}$ the corresponding dimensions are
\begin{equation}
\Delta_{m,n}^{\text{(NS)}}=\left.  \Delta_{\text{NS}}(a)\right|  _{a=a_{m,n}%
}\;\;\;\;\text{or\ \ \ \ }\Delta_{m,n}^{\text{(R)}}=\left.  \Delta_{\text{R}%
}(a)\right|  _{a=a_{m,n}}\label{Deltamn}%
\end{equation}
and one singular vector appears at level $mn/2$ in the Verma module, over
$V_{a_{m,n}}=V_{m,n}$ (at $m-n\in2Z$) or over $R_{a_{m,n}}=R_{m,n}$
($m-n\in2Z+1$) respectively. It is convenient to introduce for each pair
$(m,n)$ a ``singular vector creation operator'' $D_{m,n}$, which is graded
polynomial in $G_{-k}$ and $L_{-k}$ of level $mn/2$ with coefficients a
functions of the central charge parameter $b^{2}$, such that the singular
vector appears when $D_{m,n}$ is applied to $V_{m,n}$ or $R_{m,n}$, whichever
appropriate. In the NS case the normalization is unambiguously fixed through
the coefficient near the highest order term $D_{m,n}=G_{-1/2}^{mn}+\ldots$.
Apparently the fermion parity of this operator is that of the product $mn$. In
the R case $mn$ is always even, while $G_{0}$ allows in any case to choose
$D_{m,n}$ bosonic. Let us agree to put all the fermion operators $G_{-k}$ to
the right from the bosonic ones $L_{-k}$, arranging each group in the order of
increasing index $-k$. Then an unambiguous normalization $D_{m,n}%
=L_{-1}^{mn/2}+\ldots$ is prescribed by the coefficient near $L_{-1}^{mn/2}$.

At moderate $(m,n)$ the polynomial $D_{m,n}$ can be carried out manually. Here
we list few first examples.

\begin{itemize}
\item  Level 1/2: A singular module is over $V_{1,1}=V_{0}$ with the singular
vector crated by
\begin{equation}
D_{1,1}=G_{-1/2}\label{D11}%
\end{equation}

\item  Level 1: A singular vector in the module over $R_{1,2}$ with
\begin{equation}
D_{1,2}=L_{-1}-\frac{2b^{2}}{1+2b^{2}}G_{-1}G_{0}\label{D12}%
\end{equation}
appears at $a=a_{1,2}=-b/2$. There is similar singular module over $R_{2,1}$.
It is however needless to write down $D_{2,1}$ separately, as it is obtained
from $D_{1,2}$ through the symmetry $m\leftrightarrow n$, $b\rightarrow
b^{-1}$. Henceforth we will systematically omit such mirror images without any
special warnings.

\item  Level 3/2: There is an NS singular vector over $V_{1,3}$ with
\begin{equation}
D_{1,3}=L_{-1}G_{-1/2}+b^{2}G_{-3/2}\label{D13}%
\end{equation}

\item  Level 2: A singular representation over $R_{1,4}$ with the creation
operator
\begin{equation}
D_{1,4}=L_{-1}^{2}+\frac{3b^{2}}2L_{-2}+\frac{b^{2}(1-6b^{2})}{1+4b^{2}}%
G_{-2}G_{0}-\frac{4b^{2}}{1+4b^{2}}L_{-1}G_{-1}G_{0}\;,\label{D14}%
\end{equation}
and yet another degenerate representation of NS type, where
\begin{equation}
D_{2,2}=L_{-1}^{2}+\frac{(1+b^{2})^{2}}{2b^{2}}L_{-2}-G_{-3/2}G_{-1/2}%
\label{D22}%
\end{equation}

\item  Level 5/2: An NS degenerate field $V_{1,5}$ with
\begin{equation}
D_{1,5}=L_{-1}^{2}G_{-1/2}+2b^{2}(1+3b^{2})G_{-5/2}+3b^{2}G_{-3/2}%
L_{-1}+2b^{2}L_{-2}G_{-1/2}\label{D15}%
\end{equation}

\item  Level 3: Here we have two Ramond degenerate moduli over $R_{1,6}$ and
$R_{2,3}$, the corresponding creation operators being
\begin{align}
D_{1,6}  & =L_{-1}^{3}+\frac{13b^{2}}2L_{-2}L_{-1}+\left(  3b^{2}%
+10b^{4}\right)  L_{-3}\nonumber\\
& \ \ -\frac{6b^{2}}{1+6\,b^{2}}L_{-1}^{2}G_{-1}G_{0}-\frac{15b^{4}%
}{1+6\,b^{2}}L_{-2}G_{-1}G_{0}\label{D16}\\
& \ \ +\frac{3b^{2}\left(  1-8b^{2}\right)  }{1+6\,b^{2}}L_{-1}G_{-2}%
G_{0}+\frac{b^{2}}4G_{-2}G_{-1}-\frac{3\left(  b^{2}-12b^{4}+40b^{6}\right)
}{2\,\left(  1+6\,b^{2}\right)  }G_{-3}G_{0}\nonumber
\end{align}
and
\begin{align}
D_{2,3}  & =L_{-1}^{3}+\frac{1+4\,b^{4}}{2\,b^{2}}L_{-2}L_{-1}+\left(
1+3b^{2}+b^{4}\right)  L_{-3}\nonumber\\
& \ \ -\frac{1-4b^{4}}{b^{2}\left(  2+3b^{2}\right)  }L_{-2}G_{-1}%
G_{0}-\ \frac2{2+3b^{2}}L_{-1}^{2}G_{-1}G_{0}\label{D23}\\
& \ \ +\frac{1-8b^{2}}{2+3b^{2}}L_{-1}G_{-2}G_{0}+\frac{1-8b^{4}}{4b^{2}%
}G_{-2}G_{-1}-\frac{5-12b^{2}+4b^{4}}{2\left(  2+3\,b^{2}\right)  }G_{-3}%
G_{0}\nonumber
\end{align}

\item  Level 7/2: A singular vector in the seven dimensional space of this
level in the module over $V_{1,7}$ is created by
\begin{align}
D_{1,7}  & =L_{-1}^{3}G_{-1/2}+8b^{2}L_{-2}L_{-1}G_{-1/2}+\left(
4b^{2}+15b^{4}\right)  L_{-3}G_{-1/2}+6b^{2}L_{-1}^{2}G_{-3/2}\label{D17}\\
& \ \ \ \ +18b^{4}L_{-2}G_{-3/2}-2\left(  2\,b^{2}-15b^{4}\right)
L_{-1}G_{-5/2}+\left(  2b^{2}-27b^{4}+90b^{6}\right)  G_{-7/2}\nonumber
\end{align}

\item  Level 4: In the Ramond module $R_{1,8}$ the singular vector is created
by the operator
\begin{align}
D_{1,8}  & =L_{-1}^{4}+17b^{2}L_{-2}L_{-1}^{2}+2\left(  8b^{2}+33b^{4}\right)
L_{-3}L_{-1}+\frac{105b^{4}}4L_{-2}^{2}\nonumber\\
& \ \ +\frac{6b^{2}\left(  1-10b^{2}\right)  }{1+8b^{2}}L_{-1}^{2}G_{-2}%
G_{0}\ +\frac{71b^{2}+492b^{4}+1260b^{6}}8L_{-4}\nonumber\\
& \ \ -\frac{8b^{2}}{1+8b^{2}}L_{-1}^{3}G_{-1}G_{0}-\frac{76b^{4}}{1+8b^{2}%
}L_{-2}L_{-1}G_{-1}G_{0}-\frac{2\left(  19b^{4}+84b^{6}\right)  }{1+8b^{2}%
}L_{-3}G_{-1}G_{0}\label{D18}\\
& \ \ +\frac{19b^{4}-210b^{6}}{1+8b^{2}}L_{-2}G_{-2}G_{0}+b^{2}L_{-1}%
G_{-2}G_{-1}-\frac{6\left(  b^{2}-15b^{4}+60b^{6}\right)  }{1+8\,b^{2}}%
L_{-1}G_{-3}G_{0}\nonumber\\
& \ \ \ \ \ \;\;\;\;\;-\frac{3b^{2}\left(  1-6b^{2}\right)  }4G_{-3}%
G_{-1}+\frac{15b^{2}-292b^{4}+2052b^{6}-5040b^{8}}{4\left(  1+8b^{2}\right)
}G_{-4}G_{0}\nonumber
\end{align}
There is snother, NS module over $V_{2,4}$, singular at this level with
\begin{align}
D_{2,4}  & =L_{-1}^{4}+\frac{1+2b^{2}+5b^{4}}{b^{2}}L_{-2}L_{-1}^{2}%
+\frac{\left(  1+2b^{2}\right)  \left(  1+5b^{2}+3b^{4}\right)  }{b^{2}}%
L_{-3}L_{-1}\nonumber\\
& \ \ +\frac{\left(  1-b^{2}\right)  ^{2}\left(  1+3b^{2}\right)  ^{2}%
}{4\,b^{4}}L_{-2}^{2}+\frac{\left(  1+3b^{2}\right)  \,\left(  2+8b^{2}%
+11b^{4}+3b^{6}\right)  }{2\,b^{2}}L_{-4}\label{D24}\\
& \ \ -2L_{-1}^{2}G_{-3/2}G_{-1/2}+2\,\left(  1-4b^{2}\right)  L_{-1}%
G_{-5/2}G_{-1/2}-\frac{1-3\,b^{4}}{b^{2}}L_{-2}G_{-3/2}G_{-1/2}\nonumber\\
& \ \ +\frac{\left(  1+3\,b^{2}\right)  \,\left(  1+3\,b^{2}-12\,b^{4}\right)
}{4\,b^{2}}G_{-5/2}G_{-3/2}+\frac{1-14\,b^{2}+29\,b^{4}-12\,b^{6}}{4\,b^{2}%
}G_{-7/2}G_{-1/2}\nonumber
\end{align}

\item  Level 9/2: At this level there are two NS highest weight vectors,
$V_{1,9}$ and $V_{3,3}$, the corresponding $D$-operators reading
\begin{align}
D_{1,9}  & =L_{-1}^{4}G_{-1/2}+20b^{2}L_{-2}L_{-1}^{2}G_{-1/2}+b^{2}\left(
20+91b^{2}\right)  L_{-3}L_{-1}G_{-1/2}+\nonumber\\
& \ +6b^{2}\left(  2+15b^{2}+42b^{4}\right)  L_{-4}G_{-1/2}+36b^{4}L_{-2}%
^{2}G_{-1/2}+10b^{2}L_{-1}^{3}G_{-3/2}+\label{D19}\\
& \ +110b^{4}L_{-2}L_{-1}G_{-3/2}+56\left(  b^{4}+5b^{6}\right)
L_{-3}G_{-3/2}-10\,b^{2}\left(  1-9b^{2}\right)  L_{-1}^{2}G_{-5/2}%
-\nonumber\\
& \ -(38b^{4}-360b^{6})L_{-2}G_{-5/2}+2b^{2}\left(  5-81b^{2}+315b^{4}\right)
L_{-1}G_{-7/2}\nonumber\\
& \ -6b^{2}\left(  1-23b^{2}+171b^{4}-420b^{6}\right)  G_{-9/2}+b^{4}%
G_{-5/2}G_{-3/2}G_{-1/2}\nonumber
\end{align}
and
\begin{align}
\ D_{3,3}  & =L_{-1}^{4}G_{-1/2}+\frac{2\left(  1+b^{4}\right)  }{b^{2}}%
L_{-2}L_{-1}^{2}G_{-1/2}+\frac{1+2b^{2}+b^{4}+2b^{6}+b^{8}}{b^{4}}L_{-3}%
L_{-1}G_{-1/2}\nonumber\\
& \ +\frac{2\,\left(  2+5b^{2}+2b^{4}\right)  }{b^{2}}L_{-4}G_{-1/2}%
+4L_{-2}^{2}G_{-1/2}+\frac{2\left(  1-3b^{4}+b^{8}\right)  }{b^{4}}%
L_{-2}L_{-1}G_{-3/2}\label{D33}\\
& \ +\frac{1+b^{4}}{b^{2}}L_{-1}^{3}G_{-3/2}+\frac{\left(  1+b^{2}\right)
^{2}(1-6b^{4}+b^{8})}{b^{6}}L_{-3}G_{-3/2}-\frac{1-10b^{2}+b^{4}}{b^{2}}%
L_{-1}^{2}G_{-5/2}\nonumber\\
& \ +\frac{2\left(  4-9b^{2}+4b^{4}\right)  }{b^{2}}L_{-1}G_{-7/2}%
-\frac{2\left(  1-2b^{2}-7b^{4}-2b^{6}+b^{8}\right)  }{b^{4}}L_{-2}%
G_{-5/2}\nonumber\\
& \ +\frac{2\left(  2-6b^{2}+5b^{4}-6b^{6}+2b^{8}\right)  }{b^{4}}%
G_{-9/2}+\frac{1-9\,b^{4}+b^{8}}{b^{4}}G_{-5/2}G_{-3/2}G_{-1/2}\nonumber
\end{align}
\end{itemize}

We quote here these bulky expressions simply to give a general idea of what
we're dealing with. Of course the same expressions with $G_{n}\rightarrow\bar
G_{n}$ and $L_{n}\rightarrow\bar L_{n}$ give the ``left'' creation operators
$\bar D_{m,n}$.

In SLFT, like in the bosonic LFT, all singular vectors ``decouple'', i.e., in
the sense of quantum operators
\begin{equation}
D_{m,n}(V,R)_{m,n}=\bar D_{m,n}(V,R)_{m,n}=0\label{nv}%
\end{equation}
where for the primary field stands either NS or Ramond one, dependent in an
obvious way on the parity of $m$ and $n$\footnote{Below we often use this
abbreviation $(V,R)$, which stands either for $V$ or for $R$, dependent in an
obvious manner on the context.}. This is the quantum version of the classical
relations (\ref{cnull}). Precisely as in the case of LFT, equations (\ref{nv})
can be considered as the basic dynamic principle of SLFT. Even the very first
steps in the study of SLFT \cite{Arvis, DHoker, Babelon}, as well as the later
achievements in constuction of a consistent theory \cite{Pogossian, Marian},
are based on this ``decoupling''.

In the next section we state certain algebraic property of the $D_{m,n}$
operators, a supersymmetric generalization of the product formula of
\cite{higher} for the ``norms of logarithmic primaries'' (defined more
precisely below).

\textbf{Norms of logarithmic primaries.} Related to every $D_{m,n}$ of the
previous section, define a ``conjugate'' operator $D_{m,n}^{\dagger}$ through
the prescriptions $G_{n}^{\dagger}=G_{-n}$ and $L_{n}^{\dagger}=L_{-n}$. Then
$D_{m,n}^{\dagger}D_{m,n}$ obviously acts invariantly at the levels. In
particular the highest weight vector $\left|  a\right\rangle $ (here we use
unified notation for the NS and Ramond primary states) of dimension $\Delta$
($\Delta=\Delta_{\text{NS}}(a)$ or $\Delta=\Delta_{\text{R}}(a)$, dependent on
the type of the reperesentaion) is an eigenvector
\begin{equation}
D_{m,n}^{\dagger}D_{m,n}\left|  a\right\rangle =d_{m,n}(a,b)\left|
a\right\rangle \label{dmn}%
\end{equation}
with certain eigenvalue $d_{m,n}(a,b)$. By definition this function is zero at
$a=a_{m,n}$ (or $a=Q-a_{m,n}$), where $\Delta_{a}$ becomes the Kac dimension
(\ref{Deltamn}) of singular representation. We're interested in the quantity
$r_{m,n}$, the coefficient in the linear term
\begin{equation}
d_{m,n}(a,b)=r_{m,n}(a-a_{m,n})+O\left(  (a-a_{m,n})^{2}\right) \label{rmn}%
\end{equation}
. This coefficient is a function of $b$. ``Manual'' calculations with the
explicit expressions (\ref{D11}--\ref{D33}) produce the following compact
results
\begin{align}
r_{1,1} &  =b^{-1}(1+b^{2})\;\;\;\nonumber\\
\ r_{1,2} &  =b^{-1}(1-b^{4})\nonumber\\
\ \ \ r_{1,3} &  =-b^{-1}(1-b^{4})(1+3b^{2})\nonumber\\
r_{1,4} &  =-2b^{-1}(1-b^{4})(1-9b^{4})\nonumber\\
r_{2,2} &  =b^{-5}(1-b^{4})^{2}(1+b^{2})^{3}\nonumber\\
r_{1,5} &  =4b^{-1}(1-b^{4})(1-9b^{4})(1+5b^{2})\label{rrmn}\\
r_{1,6} &  =12b^{-1}(1-b^{4})(1-9b^{4})(1-25b^{4})\nonumber\\
r_{2,3} &  =b^{-7}(1-b^{4})^{3}(1-9b^{4})\nonumber\\
r_{1,7} &  =-36b^{-1}(1-b^{4})(1-9b^{4})(1-25b^{4})(1+7b^{2})\nonumber\\
r_{1,8} &  =-144b^{-1}(1-b^{4})(1-9b^{4})(1-25b^{4})(1-49b^{4})\nonumber\\
r_{2,4} &  =-2b^{-9}(1-b^{4})^{3}(1-9b^{4})^{2}(1+2b^{2})\nonumber\\
r_{1,9} &  =579b^{-1}(1-b^{4})(1-9b^{4})(1-25b^{4})(1-49b^{4})(1+9b^{2}%
)\nonumber\\
r_{3,3} &  =-3b^{-13}(1-b^{4})^{4}(1+b^{2})(9-b^{4})(1-9b^{2})\nonumber
\end{align}
All of them fall into the ``product formula''
\begin{equation}
r_{m,n}=2^{1-mn}\prod_{(k,l)\in[m,n]}(kb^{-1}+lb)\;,\label{rprod}%
\end{equation}
where symbol $[m,n]$ denotes either $[m,n]_{\text{NS}}$, or $[m,n]_{\text{R}}
$, dependent on the type of representation. Here
\begin{align}
\lbrack m,n]_{\text{NS}} &  =\left\{  1-m:2:m-1,1-n:2:n-1\right\}
\label{NSset}\\
&  \cup\left\{  2-m:2:m,2-n:2:n\right\}  \setminus\{0,0\}\nonumber\\
\lbrack m,n]_{\text{R}} &  =\left\{  1-m:2:m-1,2-n:2:n\right\} \label{Rset}\\
&  \cup\left\{  2-m:2:m,1-n:2:n-1\right\}  \setminus\{0,0\}\nonumber
\end{align}
In these expressions $a:d:b$ (``from $a$ to $b$ step $d$'') stands for the
``linear'' set, i.e., the set of numbers $a,a+d,a+2d,\ldots,b$. Symbol
$\left\{  A,B\right\}  $ is for the set of pairs $(k,l)$ with $k$ and $l$
running independently the sets $A$ and $B$ and $\left\{  A_{1},B_{1}\right\}
\cup\left\{  A_{2},M_{2}\right\}  $ is the standard union of two sets.
Finally, $\ldots\setminus\{0,0\}$ means that the pair $(0,0)$ is excluded.

Expression (\ref{rprod}) is very much like the one obtained in \cite{higher}
for the similar characteristic related to the singular representations of the
usual Virasoro algebra. In that case a line of ``physical'' arguments has been
proposed, based on the consistency of HEM's with the one point functions in
the so called ``Poincar\'e disk geometry'' (see \cite{ZZ}). At the same time,
it is clear that the product formula is of purely algebraic nature and has
nothing to do neither with HEM's nor with the Poincar\'e disk. It is desirable
therefore to have a direct algebraic derivation of the product formula, as
well as of its SUSY version (\ref{rprod}). It is plausible that such a
derivation can be found studying the structure of moduli embeddings in the
non-trivial case of rational $b^{2}$ (authors thank B. Feigin for a discussion
of this point). Let us mention also an algebraic, althouth rather complicated
proof of a particular case of the product formula, related to the singular
representations $(1,n)$ of the Virasoro algebra \cite{Imbimbo}.

In this short note we follow a simplified way, opposite to that of
\cite{higher}. In the absence of a direct proof, we take eq.(\ref{rprod}) for
granted and derive the coefficients in HEM's comparing it with the one point
functions on the Poincar\'e disk \cite{Fukuda, Ahn}. This procedure, unlike
the study of multipoint functions in \cite{higher}, makes the analysis very
compact and allows to avoid heavy calculations with the SLFT stucture constants.

\textbf{Logarithmic degenerate fields and HEM's.} Now we're in the position to
define, in the spirit of ref.\cite{higher}, the set of ``logarithmic
degenerate fields'' $V_{m,n}^{\prime}$ ($m-n\in2Z$) and $R_{m,n}^{\prime}$
($m-n\in2Z+1$). General logarithmic fields $V_{a}^{\prime}=\partial
V_{a}/\partial a$ and $R_{a}^{\prime}=\partial R_{a}/\partial a$ are the
derivatives in $a$ (a normal ordered free fields $:\phi\exp(a\phi):$ and
$\sigma(\mu):\phi\exp(a\phi):$ is what they look like in the $\phi
\rightarrow-\infty$ free field limit) of the corresponding primary ones. Let
us set
\begin{align}
V_{m,n}^{\prime} &  =\left.  V_{a}^{\prime}\right|  _{a=a_{m,n}}%
\;\;\;\;\;m-n\in2Z\label{VRprim}\\
R_{m,n}^{\prime} &  =\left.  R_{a}^{\prime}\right|  _{a=a_{m,n}}%
\;\;\;\;\;m-n\in2Z+1\nonumber
\end{align}
Whereas $V_{m,n}^{\prime}$ and $R_{m,n}^{\prime}$ are logarithmic fields (as
well as general $V_{a}^{\prime}$ and $R_{a}^{\prime}$), holds true the following

\textit{Proposition:}
\begin{equation}
\bar D_{m,n}D_{m,n}(V,R)_{m,n}^{\prime}\label{DDVR}%
\end{equation}
are primary fields.

We will not repeat the proof here, as it follows literally the considerations
of ref.\cite{higher}. The idea is quite simple. From the algebraic point of
view the only difference between the logarithmic fields $(V,R)_{a}^{\prime}$
and the primary ones $(V,R)_{a}$ is the inhomogeneous term in the action of
$L_{0}$
\begin{equation}
L_{0}(V,R)_{a}^{\prime}=\Delta_{\text{(NS,R)}}(a)(V,R)_{a}^{\prime}%
+\frac{d\Delta_{\text{(NS,R)}}(a)}{da}(V,R)_{a}\label{Log}%
\end{equation}
and the same for $\bar L_{0}$. Hence, for avery $k>0$ the vectors
$L_{k}D_{m,n}(V,R)_{m,n}^{\prime}$ and $G_{k}D_{m,n}(V,R)_{a}^{\prime}$ are in
the $SV$ (``right'') module over $(V,R)_{a}$, and therefore are annihilated by
$\bar D_{m,n}$.

Similarly to the Liouville case, in SLFT the primary fields (\ref{DDVR}) are
to be identified with other exponential fields. Comparing dimensions we find
\begin{equation}
\bar D_{m,n}D_{m,n}(V,R)_{m,n}^{\prime}=B_{m,n}(\tilde V,\tilde R)_{m,n}%
\label{HEM}%
\end{equation}
Here the exponential primaries $\tilde V_{m,n}=V_{a_{m,-n}}$ and $\tilde
R_{m,n}=R_{a_{m,-n}}$ have dimensions $\Delta_{m,n}^{\text{(NS)}}+mn/2$ and
$\Delta_{m,n}^{\text{(R)}}+mn/2$ respectively.

Equations (\ref{HEM}) are our long anticipated SLFT HEM's. The remaining
problem of the numerical coefficients $B_{m,n}$ is discussed in the next
section. We will do this comparing right and left hand sides of (\ref{HEM})
inside correlation functions. The simplest one is the one-point function in
the non-compact geometry of the Poincar\'e disk. In this geometry, unlike
sphere or ``finite disk'' \cite{FZZ}, the gauge group $SL(2,R)$ is an isometry
and therefore there is no problem of factoring out its orbits. There are thus
all reasons to expect the operator-valued relations to hold already on the one
point function level. On the other hand, such one-point functions are
relatively simple \cite{Fukuda, Ahn}. Let us turn to their discussion.

\textbf{One point functions on the Poincar\'e disk.} In refs.\cite{Fukuda,
Ahn} the one point functions were constructed in the so called Poincar\'e disk
geometry. Roughly speaking, this geometry is a quantum version of the
``basic'' classical solution to the classical SLFT equations of motion
(\ref{csliouv}) inside the unit disk $\left|  z\right|  <1$
\begin{equation}
e^{\varphi}=\frac2{M^{2}(1-z\bar z)^{2}}\ ;\ \ \ \psi=\bar\psi
=0\;.\label{basic}%
\end{equation}
The object of study is the one point funcitons of the exponential SLFT fields
(the meaning of the index $(m,n)$ near the one point functions, boundary
states and amplitudes is explained in \cite{ZZ} and \cite{Fukuda, Ahn})
\begin{equation}
\left\langle (V,R)_{a}\right\rangle _{(1,1)}=\frac{\left\langle \text{B}%
_{(1,1)}|(V,R)_{a}\right\rangle }{\left\langle \text{B}_{(1,1)}|V_{0}%
\right\rangle }=U_{(1,1)}^{\text{(NS,R)}}(a)\;,\label{onepont}%
\end{equation}
where we denoted as $\left\langle \text{B}_{(1,1)}\right|  $ the boundary
state radiated by the absolute of the Poincar\'e disk. In this paper we will
not repeat the considerations of refs.\cite{Fukuda, Ahn}, quoting only the net
result
\begin{align}
U_{(1,1)}^{\text{(NS)}}(a) &  =\frac{\left[  \pi\mu\gamma\left(  \dfrac
{Qb}2\right)  \right]  ^{-a/b}\Gamma\left(  \dfrac{Qb}2\right)  \Gamma\left(
\dfrac Q{2b}\right)  Q}{2(Q-2a)\Gamma\left(  \dfrac{Qb}2-ab\right)
\Gamma\left(  \dfrac Q{2b}-\dfrac ab\right)  }\label{Gamma}\\
U_{(1,1)}^{\text{(R)}}(a) &  =\frac{\left[  \pi\mu\gamma\left(  \dfrac
{Qb}2\right)  \right]  ^{-a/b}\Gamma\left(  \dfrac{Qb}2\right)  \Gamma\left(
\dfrac Q{2b}\right)  Q}{2\Gamma\left(  \dfrac12+\dfrac{Qb}2-ab\right)
\Gamma\left(  \dfrac12+\dfrac Q{2b}-\dfrac ab\right)  }\nonumber
\end{align}
Substituting equations (\ref{HEM}) to these one point functions, we obtain
\begin{equation}
\left\langle \text{B}_{(1,1)}|\bar D_{m,n}D_{m,n}(V,R)_{m,n}^{\prime
}\right\rangle =B_{m,n}\left\langle \text{B}_{(1,1)}|(\tilde V,\tilde
R)_{m,n}\right\rangle \label{X}%
\end{equation}
The boundary state $\left\langle \text{B}_{(1,1)}\right|  $ enjoys
superconformal invariance. This means that for all $n\in Z$ and all $k\in
Z,Z+1/2$ the following identities hold
\begin{align}
\left\langle \text{B}_{(1,1)}\right|  \bar G_{k} &  =i\left\langle
\text{B}_{(1,1)}\right|  G_{-k}\label{scbound}\\
\left\langle \text{B}_{(1,1))}\right|  \bar L_{n} &  =\ \left\langle
\text{B}_{(1,1)}\right|  L_{-n}\;.\nonumber
\end{align}
It is easy to see that these identities entail (operator $D_{m,n}^{\dagger}$
is defined in sect. 5)
\begin{align}
\left\langle \text{B}_{(1,1)}\right|  \bar D_{m,n} &  =\,\left\langle
\text{B}_{(1,1)}\right|  D_{m,n}^{\dagger}\ \ \ \ \;\ \;mn\in2Z\label{DDagger}%
\\
\left\langle \text{B}_{(1,1)}\right|  \bar D_{m,n} &  =i\left\langle
\text{B}_{(1,1)}\right|  D_{m,n}^{\dagger}\ \ \ \ \ \ mn\in2Z+1\nonumber
\end{align}
In the last case a multiplier $i$ remains because at $mn\in2Z+1$, unlike all
other cases, the singular vector creating operator is fermionic (odd in $G$).
We find
\begin{equation}
\left\langle \text{B}_{(1,1)}\right|  \left.  D_{m,n}^{\dagger}D_{m,n}%
(V,R)_{m,n}^{\prime}\right\rangle =(-i)^{mn-2[mn/2]}B_{m,n}\left\langle
\text{B}_{(1,1)}|(\tilde V,\tilde R)_{m,n}\right\rangle \;.\label{DDV}%
\end{equation}
In terms on the one point functions eq.(\ref{rmn}) has the following
interpretation
\begin{equation}
\left\langle \text{B}_{(1,1)}|D_{m,n}^{\dagger}D_{m,n}(V,R)_{m,n}^{\prime
}\right\rangle =r_{m,n}\left\langle \text{B}_{(1,1)}|(V,R)_{m,n}\right\rangle
\label{DDr}%
\end{equation}
where numbers $r_{m,n}$ are from the product formulas (\ref{rprod}%
),(\ref{NSset}) and (\ref{Rset}). Therefore
\begin{equation}
B_{m,n}=i^{mn-2[mn/2]}\frac{r_{m,n}U_{(1,1)}^{\text{(NS,R)}}(a_{m,n}%
)}{U_{(1,1)}^{\text{(NS,R)}}(a_{m,-n})}\label{BrU}%
\end{equation}
This equality allows to find easily the coefficients $B_{m,n}$. Two cases
should be distinguished

1. NS case ($m$ and $n$ both either even or odd)
\begin{equation}
B_{m,n}=2^{mn}i^{mn-2[mn/2]}b^{n-m+1}\left[  \pi\mu\gamma(bQ/2)\right]
^{n}\gamma\left(  \frac{m-nb^{2}}2\right)  \prod\nolimits_{(k,l)\in
\left\langle m,n\right\rangle _{\text{NS}}}\lambda_{k,l}\label{oo}%
\end{equation}

2. R case ($m$ odd and $n$ even)
\begin{equation}
B_{m,n}=2^{mn}b^{n-m}\left[  \pi\mu\gamma(bQ/2)\right]  ^{n}\gamma\left(
\frac12+\frac{m-nb^{2}}2\right)  \prod\nolimits_{(k,l)\in\left\langle
m,n\right\rangle _{\text{R}}}\lambda_{k,l}\label{oe}%
\end{equation}
Symbol $\lambda_{k,l}$ is defined in (\ref{lmn}) while the sets $\left\langle
m,n\right\rangle _{\text{NS}}$ and $\left\langle m,n\right\rangle _{\text{R}}$
include the following pairs of integers $(k,l)$
\begin{align}
&  \left\langle m,n\right\rangle _{\text{NS}}=\left\{
1-m:2:m-1,1-n:2:n-1\right\} \label{oeset}\\
&  \cup\left\{  2-m:2:m-2,2-n:2:n-2\right\}  \setminus\{0,0\}\nonumber\\
&  \left\langle m,n\right\rangle _{\text{R}}=\left\{
1-m:2:m-1,2-n:2:n-2\right\} \label{eoset}\\
&  \cup\left\{  2-m:2:m-2,1-n:2:n-1\right\}  \setminus\{0,0\}\nonumber
\end{align}

Only the series $(1,2k-1)$, $k=1,2,\ldots$ of HEM's allows a classical limit.
It is easy to check that (\ref{oo}) at $b\rightarrow0$ turns to $B_{1,2k-1}%
\rightarrow i(-)^{k-1}b^{-1}(2\pi\mu b^{2})^{2k-1}[(k-1)!]^{2}$, in agreement
with the results (\ref{todo}) of the classical calculations.

\textbf{Acknowledgements.} Collaboration of the authors was possible in the
framework of the INTAS project, grant INTAS-OPEN-03-51-3350 2004-2005.
Research work of A.B. was supported by grants RFBR 04-02-16027 2003-2005 and
SS-2044-2003.2, and also by the RAS program ''Elementary particles and
Fundamental nuclear physics''. The study was finished while Al.Z visited RIMS
and Yukawa Institute, University of Kyoto. He thanks these Institutes, and
especially T.Miwa and R.Sasaki for support and stimulating discussions. In
addition this visit was supported by the EGIDE project. Al.Z was also
sponsored by the European Committee under contract EUCLID HRPN-CT-2002-00325.

\end{document}